\documentclass[useams,usenatbib,a4]{mn2e}

\usepackage {multirow}
\usepackage {graphicx}
\usepackage[hyperindex,breaklinks=true, colorlinks, citecolor=blue]{hyperref}

\newcommand{\imgspace}{.47\textwidth}

\newcommand{\celsius} {\ensuremath{^{\circ}}}
\newcommand{\angdeg} {\ensuremath{^{\circ}}}
\newcommand{\angmin} {\ensuremath{'}}

\title[Radio Determination of the New Moon]{A Radio Determination of the Time of the New Moon}
\author[Y. A. Hafez et al.]{Yaser A. Hafez$^1$,
  Lorenzo Trojan$^2$,
  Fahad H. Albaqami$^1$,
  \newauthor Abdulmajeed Z. Almutairi$^1$,
  Rodney D. Davies$^2$,
  Clive Dickinson$^2$,
  Lucio Piccirillo$^2$\\
$^1$ National Center for Mathematics and Physics, KACST, P.O. Box 6086, Riyadh, 11442, Saudi Arabia \\
$^2$ Jodrell Bank Centre for Astrophysics, Alan Turing Building, School of Physics \& Astronomy, \\
The University of Manchester, Oxford Road, Manchester, M13 9PL 
}

\begin{document}

\maketitle

\begin{abstract}

The detection of the New Moon at sunset is of importance to
communities based on the lunar calendar. This is traditionally
undertaken with visual observations. We propose a radio method which
allows a higher visibility of the Moon relative to the Sun and
consequently gives us the ability to detect the Moon much closer to
the
Sun than is the case of visual observation. We first compare the
relative brightness of the Moon and Sun over a range of possible
frequencies and find the range 5--100\,GHz to be suitable. The next
consideration is the atmospheric absorption/emission due to water
vapour and oxygen as a function of frequency. This is particularly
important since the relevant observations are near the horizon.
We show that a frequency of $\sim 10$ GHz is optimal for this programme.
We have designed and constructed a telescope with a FWHM resolution of
0$^\circ{}\!\!$.6 and low sidelobes to demonstrate the potential of this
approach. At the time of the 21 May 2012 New Moon the Sun/Moon
brightness temperature ratio was $72.7 \pm 2.2$  in agreement with
predictions
from the literature when combined with the observed sunspot
numbers for the day. The Moon would have been readily detectable at
$\sim 2^{\circ}$ from the Sun. Our
observations at 16\,hr\,36\,min UT indicated that the Moon
would have been at closest approach to the Sun
16\,hr\,25\,min earlier; this was the annular solar eclipse of
00\,hr\,00\,min\,UT on 21 May 2012.
\end{abstract}


\section{Introduction}
\label{sec:introduction}

Throughout history, dating back to Babylonian times, the first
appearance of the New Moon has been used to determine the
calendar. Hindu, Hebrew and Muslim calendars are based on the visual
sighting of the first cresent Moon after conjunction with the Sun at
the beginning of each month \citep{Bruin1977}. Today, most lunar
calendars are based on calculation with a
mean lunar synodic period of 29.35 solar days. However, many observe a
religious lunar calendar using actual visual observations of the
setting crescent Moon.

The visual sighting by the naked eye of the first crescent after New
Moon (Sun-Moon conjunction) depends on
the angular distance between the Moon and the
Sun at sunset. Typically, twenty four hours after conjunction, the
Moon has moved $\sim 12$\angdeg{} in Right Ascention (RA) from the Sun and the Moon
illumination will be about 1~per cent. Under these conditions, the Moon
will set approximately 40\,min after sunset
(depending on the observer's geographical latitude) when the sky
luminosity has decreased.
Provided the sky is sufficiently clear,
these factors (lower sky luminosity and and increased illumination)
combine to make the
crescent Moon visible for few minutes with the naked eye before it
sets. If sunset occurs much less
than twenty four hours after
conjunction, the angular separation between the Moon and the Sun is
not sufficient for the sighting to be performed.

We propose a radio method for the determination of
the time of New Moon. This method is independent of the weather and
can be used to establish the exact time of the New Moon with much
higher accuracy than the traditional visual method.
At optical frequencies, the full Moon is a factor of 4$\times$10$^{5}$
fainter than the Sun; in the radio range, this factor is reduced to
approximately 100--500.
The challenge is
to detect the weaker signal from the Moon in the presence of the much
brighter Sun and also to account for the atmospheric emission and
absorption at low elevation.
Both the Sun and the Moon have variable brightness. The Sun changes
with the 27-day solar rotation and the 11.3-year sunspot cycle. The
Moon varies with the lunar 29.3-day cycle.  In addition the Sun shows
strong variation with frequency; the Moon is less variable. The
choice of an optimal radio frequency is critical for this project.

The paper is set out as follows. Section~\ref{sec:radiosun} describes
the frequency
dependence of the various components of solar emission.
Section~\ref{sec:radiomoon}
gives the frequency dependence of the lunar emission including the
variation with lunar phase.
The
atmospheric opacity of H$_2$O and O$_2$ are discussed in
Section~\ref{sec:atmemission}. This is
required because the observations are made at low elevation where
absorption and emission effects are
significant. Section~\ref{sec:proposedsys} describes a working system
at 10 GHz. It includes
observations 16\,hr after New Moon on 21 May 2012, demonstrating the
potential of such a system for direct observation of the New Moon.
Conclusions are given in Section~\ref{sec:conclusion}.

\section{The radio properties of the Sun}
\label{sec:radiosun}

The radio emission of the Sun comes from the corona, the chromosphere
and the photosphere in various proportions depending on the frequency.
The $10^{6}$\,K corona dominates the emission at frequencies below
$\sim 1$\,GHz where its free-free emission becomes optically thick.
At 50\,GHz and above the photosphere (electron temperature, $T_e = 6000$\,K) is the major
contributor.  At 5 to 30\,GHz the chromosphere ($T_e = 10000$\,K) is
also a significant component and is the source of the radio
variability.

We now consider the relationship between the optical and radio
emission from the Sun in its quiet and active modes.

\subsection{The brightness temperature of the quiet Sun}

The brightness temperature, $T_b$, of the Sun at a given frequency can
be expressed as the value which will give the observed flux density
when averaged over the optical diameter.  We confine our study to
frequencies above 1\,GHz where $T_b < 10^{5}$\,K in order that the Sun
is considerably less than 1000 times the Moon emission (see
Section~\ref{sec:radiomoon}).  The quiet Sun emission is defined as that
observed/estimated for zero sunspots.  The quiet Sun $T_b$ values at
times of sunspot maximum and minimum are given in Table~\ref{tab:sun}
for representative frequencies of interest for this project.  These
values are derived from the multiplicative terms given by
\citet{Cox2000} and cross-checked  with data given by
\citet{Kundu1965} and by \citet{Covington1954}.

Table~\ref{tab:sun} shows that for frequencies less than 10\,GHz the
quiet Sun $T_b$ increases approximately as wavelength squared, as expected for an
optically thin solar corona. At 10--20\,GHz the emission is dominated
by the optically thick chromosphere; at 50\,GHz and above we approach
the optically thick photosphere with $T_b = 6000$\,K.  The ratio of
the quiet Sun $T_b$ at sunspot maximum to minimum is highest ($\sim
1.5$) at frequencies dominated by coronal emission.  At 10\,GHz the
ratio is 1.32.

\begin{table}
  \centering
  \begin{tabular}{cc|cc|cc}
    \hline
    &      &\multicolumn{2}{|c|}{Quiet Sun $T_b$ }   & $T_b$ ratio & Increase of  \\
    $\lambda$ &$\nu$ &SS min    &SS max              &             & SS min $T_b$ \\
    (cm)  & (GHz)&(1000\,K) & (1000\,K)              &             & (per cent)   \\
    (1) & (2) & (3) & (4) & (5) & (6) \\
    \hline \hline
    30            &1.0   &132       &218            &1.65         & 110         \\
    15            &2.0   &56        &85             &1.51         & 126         \\
    \hline
    6.0           &5.0   &21.4      &31.6           &1.48         & 68          \\
    3.0           &10    &12.6      &16.6           &1.32         & 17          \\
    1.5           &20    &9.5       &11.0           &1.16         & 2           \\
    \hline
    0.6           &50    &6.8       &7.4            &1.10         & 0           \\
    0.3           &100   &6.3       &6.5            &1.03         & 0           \\
    \hline
  \end{tabular}
  \caption{
    The radio brightness temperature of the Sun in the frequency
    range 1--100\,GHz.  The sunspot maximum (SS max) and minimum (SS min) values
    for the steady
    component (i.e. at $R = 0$) are given in columns 3 and 4.  Column 5
    gives the brightness temperature ratio between sunspot maximum and
    minimum. Column 6 gives the component correlated with sunspot number
    for $R = 100$. $R$ is defined as $10N_g + N_s$ where $N_g$ is the number
    of sunspot groups and $N_s$ is the number of individual
    sunspots.
  }
  \label{tab:sun}
\end{table}

\subsection{Variation of emission over the sunspot cycle}

The 11.3-year sunspot cycle can be seen at all frequencies from radio
to X-rays. In the frequency range of 1--100\,GHz considered here the
emission is closely correlated with the sunspot activity measured
either as the sunspot area or the sunspot number.  It has become
conventional to use the sunspot number $R$, defined as $R = 10N_g +
N_s$ where $N_g$ is the number of sunspot groups and $N_s$ is the
number of individual spots \citep{Cox2000}.

\begin{figure}
  \centering
  \includegraphics[width=\imgspace]{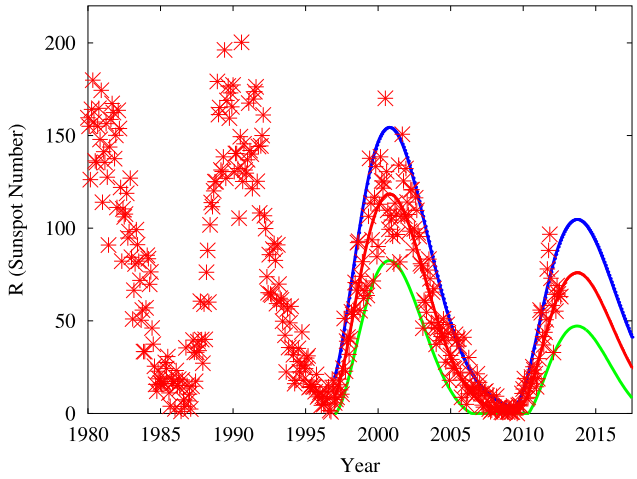}
  \caption{
    Monthly sunspot numbers ($R$) from 1980 to the
    present. The sunspot
    number data are from the Solar Influences Data Analysis
    Center.$^{1}$ The solar cycle model for the
    23$^{rd}$ cycle (1998--2009) and the prediction for the 24$^{th}$
    cycle (2009--2020) in red line is based on
    \citet{Hathaway1994}. The blue line gives the prediction for
    95\,per cent of the observations being below this level and the
    green line is for 95\,per cent above.
  }
  \label{fig:sunspotmonth}
\end{figure}

Fig.~\ref{fig:sunspotmonth} shows the
monthly averaged values of $R$ from 1980 to the present.  Predicted
values for the remainder of the current sunspot maximum are shown as
full lines to represent the mean value of $R$ and the upper and lower
bounds. The sunspot data are from the Solar Influences Data Analysis
Center
\footnote{The Solar
Influences Data Analysis Center website http://sidc.oma.be/index.php}.
It
is seen that the level of activity varies from cycle to cycle.  For
example the most recent cycle lasted for an unusually long period and
the present maximum promises to be weaker than average.

\begin{figure}
  \centering
  \includegraphics[width=\imgspace]{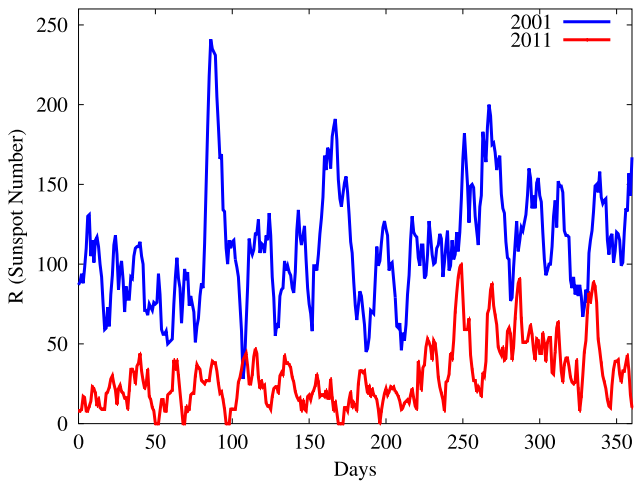}
  \caption{
    Daily sunspot numbers, $R$, covering a year near the
    previous sunspot maximum (2001) in blue compared with a year at
    the rise
    from the recent sunspot minimum (2011) plotted in red.  The 27-day
    period of the longer-lasting groups can be seen in both years. Note
    the ordinate is sunspot number $R$. Data are taken from the
    published values on the Solar Influences Data Analysis Center
    website.$^{1}$
  }
  \label{fig:sunspotday}
\end{figure}

A shorter period of variability arises from the 27-day rotation period
of the Sun as defined by the active areas with their associated
sunspots.  These active areas can last from a few days to a few months
as the sunspots are born and decay.  The larger sunspot groups last
for several rotation periods as can be seen in
Fig.~\ref{fig:sunspotday}, which shows the daily values of $R$ for 2001
(near sunspot maximum) and for 2011 (near the most recent sunspot
minimum).

\begin{figure}
  \centering
  \includegraphics[width=\imgspace]{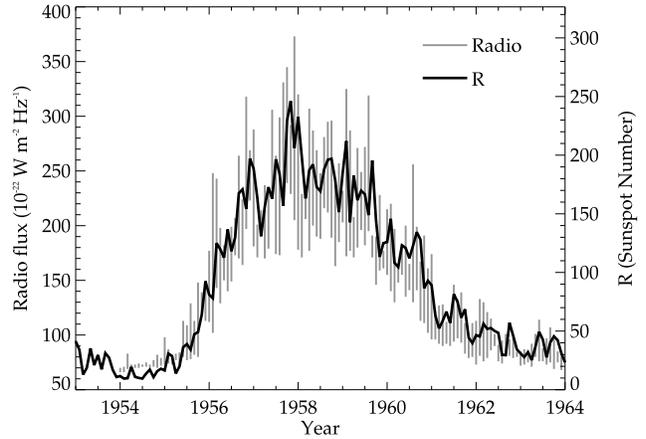}
  \caption{
    A comparison of the 2.8\,GHz flux density in units of
    $10^{-22}$\,W\,m$^{-2}$\,Hz$^{-1}$ with the sunspot number, $R$, for
    the period of one sunspot cycle from 1954 to 1964.
    The range of flux
    density in each month is plotted.
    Note that the radio
    emission for R$\sim$0 (sunspot minimum) corresponds to $\sim 60$ flux
    units due to the steady chromosphere emission
    The radio data are from \citet{Covington1974}; the sunspot data are
    from the Solar Influences Data Analysis Center$^{1}$.
  }
  \label{fig:opticalradio}
\end{figure}

The close correlation between the 2.8\,GHz radio emission
\citep{Covington1974} and the monthly values of sunspot number is
illustrated in Fig.~\ref{fig:opticalradio} which covers the sunspot
cycle running from 1954 to 1964.   The quiet Sun at sunspot minimum is
66 flux units and rises to $\sim 120$ flux units at sunspot
maximum. The variable $R$-correlated emission more than doubles this
level. This result can be checked against the predictions of
Table~\ref{tab:sun}.  The peak sunspot number is $R = 250$ which would
add another 120 flux units at sunspot maximum giving a total of $\sim
240$ flux units in agreement with Fig.~\ref{fig:opticalradio}. Clearly
there is significant variation in the radio emission over the solar
cycle that needs to be taken into account in the present study.

\subsection{Variability over timescales $<<1$ day}

The main variability on timescales $<< 1$ day arises from the radio bursts
associated with solar flares.  At 2.8\,GHz, for
example, typical bursts last 2\,min to 50\,min \citep{Covington1954}.  These occur in various
forms, with the two most common being a single burst and a single burst
with a following enhancement.  Peak flux values are 10--20 per cent of
the quiet Sun.  Analysis of the flare data
of \citet{Dodson1954}
indicated that 57 per cent of the brightest flares (types 2 and 3) had
associated radio bursts.  The frequency of occurrence of the stronger
bursts was one every few days and consequently will have little effect
on the present programme, which tracks the Sun/Moon for periods of
several hours at least.

\subsection{The diameter of the quiet and active Sun as a function of frequency}

This is a small effect at frequencies where the chromospheric and
photospheric emission dominates, being the optical value.  The solar
optical diameter at the mean Sun-Earth distance is 32.0\,arcmin. The
diameter
changes by $\pm 1.67$ per cent beween aphelion and perihelion.  Consequently
at 10\,GHz and above the diameter can be taken as the optical value but
with some limb-darkening in the polar direction.

At 1.4 GHz the diameter of the quiet Sun is 1.1 times the optical
value \citep{Christiansen1955}. At the much lower frequency
of 200\,MHz the effective diameter is $1.5 \times
1.2$ times the optical value
with the greater extent in the equatorial direction
\citep{Kundu1965}.

The active Sun at frequencies above 1\,GHz is produced by heated
regions lying above sunspot and plage areas \citep{Christiansen1957}.
These can extend up to 5~arcmin (0.2 solar diameter) above the quiet
Sun level.

\section{The radio properties of the Moon}
\label{sec:radiomoon}

In this section we consider the properties of the Moon as a function
of frequency.  In the present case we are concerned with the detection
of the Moon in the presence of the much stronger Sun where we use an
observing beam which encompasses the whole Moon disk. The integrated
properties are the total flux density or the corresponding brightness
temperature averaged over the the lunar disc.  There is a significant
phase effect, which varies significantly in amplitude and phase over the
radio range.

\begin{table}
  \centering
  \begin{tabular}{c c c c c}
\hline
$\lambda$&$\nu$  &$T_0$          &$T_1$               & Phase from   \\
         &       &(average)      &(1$^{st}$ harmonic) &  New Moon    \\
(cm)     &(GHz)  & (K)           & (K)                & (\angdeg)    \\
\hline \hline
0.1      &300    &203            &101              &5                 \\
0.2      &150    &208            &80               &14                 \\
0.3      &100    &210            &72               &17                \\
0.4      &75     &211            &62               &24                \\
0.8      &37     &214            &38               &32                 \\
1.6      &19     &215            &29               &35                 \\
3.2      &9.4    &217            &14               &40                \\
9.6      &3.1    &221            &4                &42                 \\
20       &1.5    &224            &0.0              &--                 \\
30       &1.0    &226            &0.0              &--                 \\
\hline
  \end{tabular}
  \caption{
    Brightness temperature of the Moon at selected frequencies;
    the data are from \citet{Krotikov1987}.  Listed are the mean
    brightness
    temperatures over a lunation, $T_0$, the amplitude of the 29.3-day first
    harmonic, $T_1$, and the phase shift of this harmonic as measured from the
    time of New Moon, $\xi$.
  }
  \label{tab:moon}
\end{table}

\subsection{The mean brightness temperature at radio frequencies}

\begin{figure}
  \centering
  \includegraphics[width=\imgspace]{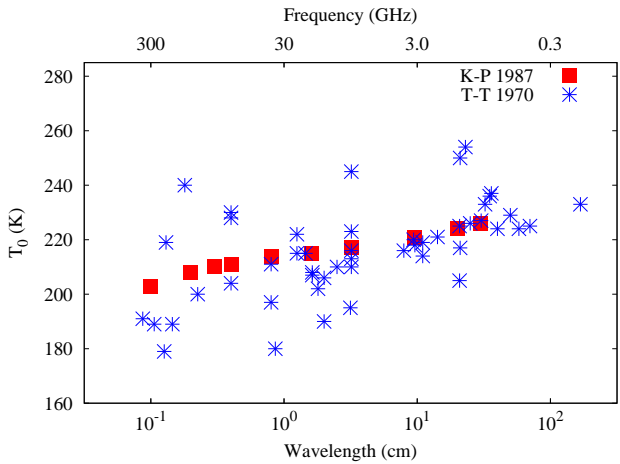}
  \caption{
    The mean brightness temperature $T_0$ of the Moon as a function
    of wavelength. The blue circle data points are from
    \citet{Troitski1970}.
    The red squares are the model of \citet{Krotikov1987}.
  }
  \label{fig:moonemissiontemp}
\end{figure}

A substantial body of observations of the Moon is available which
covers a wide frequency range \citep{Hagfors1970}.
In recent times,
the Moon has been
widely used as a calibrator for radio brightness
\citep{Linsky1973,Harrison2000,Poppi2002,quiet2012} and
lunar emission data have been used to complement
information from dust samples
returned from the Apollo landings \citep{Heiken1991} and from the
lunar orbiter CE-1 \citep{Zheng2009}. It is found that there
is a variation of brightness across the surface dependent upon the
local properties of the regolith \citep{Zhang2012}.  Theoretical
modelling of the the lunar material indicates that the depth from
which the emission comes is $\approx 10$ times the wavelength
\citep{Linsky1973}.

The mean brightness temperatures of the Moon $T_0$ taken from the
literature
\citep{Troitski1970} are shown in Fig.~\ref{fig:moonemissiontemp} for
the wavelength range 0.1 to 100\,cm.  Also plotted in red squares is a
more recent assessment of the data \citep{Krotikov1987}.  The paper by
\citet{Zhang2012} summarizes modern observations which show a mean
brightness at 1.4\,GHz ($\lambda \approx$21\,cm) of $233 \pm 2$\,K.  This
indicates that the mean temperature at a depth of $\approx 2$\,m is 20
to 30\,K higher than in the thin surface layer responsible for the
emission at sub-millimetre wavelengths
(Fig.~\ref{fig:moonemissiontemp}). Typical values of
the mean brightness temperature of the Moon are given in
Table~\ref{tab:moon} for the frequency range of concern here.

\subsection{The brightness temperature variation with lunar phase}

\begin{figure}
  \centering
  \includegraphics[width=\imgspace]{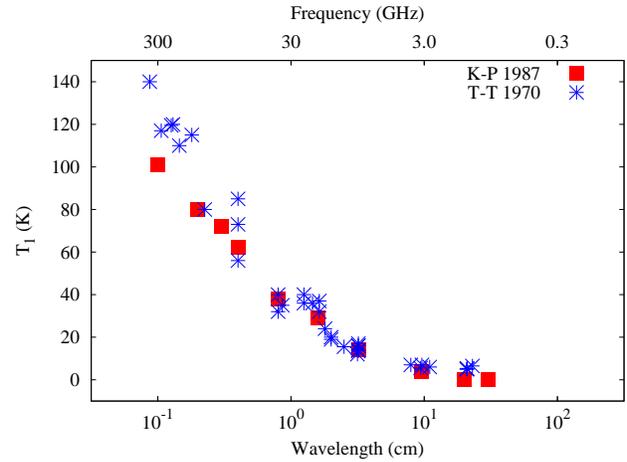}
  \caption{
    The amplitude of the first harmonic surface brightness $T_1$
    of the Moon. The description is as for Fig.~\ref{fig:moonemissiontemp}.
  }
  \label{fig:moonfirstharmonic}
\end{figure}

\begin{figure}
  \centering
  \includegraphics[width=\imgspace]{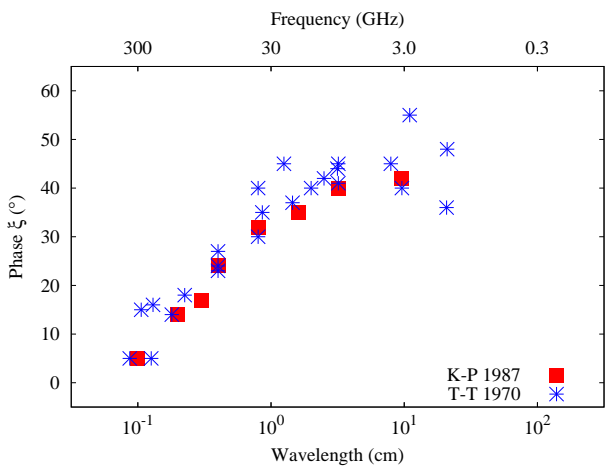}
  \caption{The phase offset of the first harmonic mean surface
  brightness measured relative to the time of New Moon. The
  description is as
  for Fig.~\ref{fig:moonemissiontemp}.}
  \label{fig:moonemissionphase}
\end{figure}

The brightness temperature of the Moon varies in frequency and time
during the 29.3-day lunar phase.  At any frequency this can be
expressed as a harmonic series.
Higher
harmonics are required at FIR and millimetre wavelengths.
The mean temperature $T_0$ and
the first
harmonic, $T_1$, are adequate to describe the temperature variation at
the lower frequencies required for the present project. We use:
\begin{equation}
{T}(\nu) = {T}_0 (\nu) + {T}_1 (\nu) \cos{}( \omega t - \xi (\nu))~,
\end{equation}
where $\omega$ is the angular frequency of the lunar cycle
(12$^{\circ}\!$.26~per~day) and $\xi (\nu)$
is the phase relative to the time of New Moon.

The amplitude of
the first harmonic phase variation is shown in
Fig.~\ref{fig:moonfirstharmonic}.  The values of $T_1$ at typical
frequencies of interest
for this project are given in Table~\ref{tab:moon}.  The source of the
data is as for Fig.~\ref{fig:moonemissiontemp} with a best-fit model
\citep{Krotikov1987} indicated by red squares.  It is seen that the
variation is greatest ($\approx 100$\,K) at high frequencies where the
emission comes from a thin surface layer. At frequencies less than
$\approx 3$\,GHz ($\lambda \ge 10$\,cm), the emission probes a depth of
$\approx 1$\,m, the
variation of $T_1$ is less than 4\,K.

The phase offset $\xi$ of the first harmonic $T_1$ as measured from
the time of New
Moon is shown in Fig.~\ref{fig:moonemissionphase} for the same data
set as used in Fig.~\ref{fig:moonemissiontemp} and
\ref{fig:moonfirstharmonic}.
Table~\ref{tab:moon}
gives values of $\xi$ at representative frequencies. 
The phase offset
is small ($<5$\angdeg{}) at frequencies $\ge 300$\,GHz ($\lambda <
0.1$\,cm)
and
rises to $\approx 40$\angdeg{} at frequencies $\le 10$\,GHz
($\lambda > 3$\,cm).
It is seen that the phase effect is significant at the frequencies of
concern for the present study.

\section{The atmospheric emission as related to the project}
\label{sec:atmemission}

The emission from the atmosphere is an important consideration when
following the Sun and the Moon to the western horizon at sunset.  The
main components that contribute to atmospheric emission at radio and
millimetre wavelengths are molecular oxygen (O$_2$) and water vapour
(H$_2$O).

The spectrum of O$_2$ at the frequencies of interest here ($<100$\,GHz) is
dominated by the pressure-broadened band of lines at $\sim 60$\,GHz.  This
emission extends through the radio band down to a few hundred MHz.  It
is the main emission component at frequencies below 10\,GHz in
relatively dry conditions where the H$_2$O spectrum is falling rapidly.

\begin{figure}
  \centering
  \includegraphics[width=\imgspace]{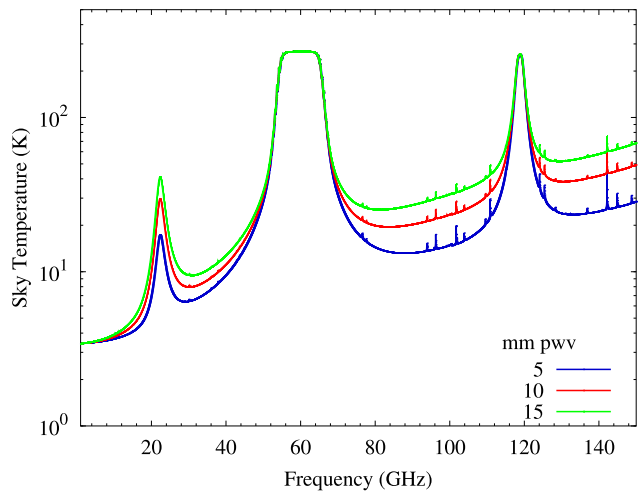}
  \caption{
    The atmospheric sky temperature at the zenith for a range of
    H$_2$O contents.
    The plots show
    the temperature for 5, 10 and
    15\,mm of precipitable water vapour and the standard sea level
    atmosphere (1013\,mbar, 288\,K). The plots are derived from the
    ATM model of \citet{Pardo2001}
  }
  \label{fig:atmemission_opacity}
\end{figure}

The components of the atmospheric emission spectrum derived from the
Atmospheric Transmission at Microwaves (ATM) model of
\citet{Pardo2001} are shown in figure \ref{fig:atmemission_opacity}
for the frequency range
1--150\,GHz. The sky brightness temperature is plotted for 5, 10 and
15 mm of precipitable water vapour (pwv) and a standard atmosphere at
sea level pressure of
1013\,mbar and a temperature of 288\,K. It is seen that the O$_2$
line is optically thick ($\tau \ge 1$) at 60\,GHz and the H$_2$O line
is optically thick at 118\,GHz. In addition to the H$_2$O lines, there
is emission from the so called ``excess water vapour absorption''
(EWA) which appears as a broad frequency component that increases with
frequency. The low opacity ranges in the total atmospheric spectrum
are at $\le 15$\,GHz, $\sim 35$\,GHz and $\sim 85$\,GHz.

Sea level values of the pwv on clear days at low ambient
temperature may be 5\,mm, but is more commonly 10--20\,mm. At heights
of 2\,km above sea level 5\,mm is more usual
\citep{Danese1989,Staggs1996}.

\begin{figure}
  \centering
  \includegraphics[width=\imgspace]{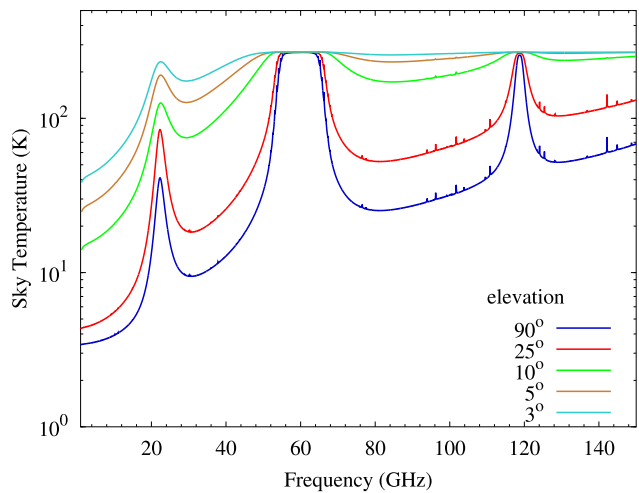}
  \caption{
    The atmospheric emission at elevations from 3\angdeg{} to
    90\angdeg{} (zenith) in the frequency range 1--150\,GHz
    the plots are derived from the ATM model of
    \citet{Pardo2001} and an H$_2$O content of 15\,mm pwv.
  }
  \label{fig:atmemission_temp}
\end{figure}

\begin{table}
  \centering
  \begin{tabular}{c c c c}
    \hline
    $\lambda$ & $\nu$     &  O$_2$ & H$_2$O       \\
    (cm)      & (GHz)     &  (K)   &  (K)         \\
    \hline                            
    \hline                            
    73        & 0.408     &  2.0   &  0.00        \\
    21        & 1.4       &  2.2   &  0.00        \\
    12        & 2.5       &  2.2   &  0.01        \\
    5.9       & 5.0       &  2.3   &  0.01        \\
    2.9       & 10        &  2.4   &  0.03        \\
    1.57      & 19        &  3.3   &  0.45        \\
    0.91      & 33        &  7.6   &  0.33        \\
    0.33      & 90        & 14.6   &  1.50        \\
    \hline
  \end{tabular}
  \caption{
    The zenith atmospheric emission from O$_2$ and H$_2$O at sea level.
    Adapted from the observational data of \citet{Danese1989},
    \citet{Staggs1996} and \citet{Zhang2012}. The O$_2$ emission is
    for the standard
    sea level atmosphere of 1013\,mbar and a temperature of
    288\,K. The H$_2$O emission per mm pwv is given.
  }
  \label{tab:atmemissionzenith}
\end{table}

Table~\ref{tab:atmemissionzenith} gives direct observational data for
the zenith emission
at frequencies of potential interest here taken from the summary by
\citet{Danese1989}, which is based on the model of
\citet{Waters1976} and \citet{Smith1982}. The frequency bands are the standard ones allocated for radio astronomy; we also include 19\,GHz, which is near the water line.
These are augmented by 1.4\,GHz data
from \citet{Staggs1996} and \citet{Zhang2012}. The O$_2$ emission is
proportional to
the local atmospheric pressure. The listed of values of the H$_2$O
emission per mm of pwv can be used along with the local water vapour
data to
give the total H$_2$O emission. Table~\ref{tab:atmemissionelevation}
gives the
total (O$_2$ + H$_2$O + EWA) zenith emission for 10\,mm pwv at
selected frequencies.

The increase in atmospheric emission at lower elevation is of
particular interest for this project. Fig.~\ref{fig:atmemission_temp}
shows
the atmospheric spectrum for elevations of 3$^{\circ}$, 5$^{\circ}$,
10$^{\circ}$, 25$^{\circ}$ and 90$^{\circ}$ (the zenith) estimated
using the ATM code. Table~\ref{tab:atmemissionelevation} lists the
total sky
temperature for 10\,mm pwv at elevations 3$^{\circ}$, 5$^{\circ}$,
10$^{\circ}$, and 90$^{\circ}$ at frequencies of interest for the
current project. For example, at 10\,GHz the sky temperature increases
from 2.6\,K at the zenith to 37\,K and 54\,K at elevations of
5$^{\circ}$ and 3$^{\circ}$ respectively. This extra emission will
significantly
increase the receiver system noise temperature at frequencies $\ge
10$\,GHz. At the higher frequencies the spatial variation of the
H$_2$O distribution may lead to unstable baselines in the recorded
data. The O$_2$ distribution is relatively steady in comparison.

In a related fashion, the corresponding optical depth of the atmosphere
at higher frequencies leads to significant absorption at lower
elevations. For example, at 10\,GHz the zenith absorption of 1~per cent
increases to 6, 12 and 20~per cent at elevations of 10$^{\circ}$,
5$^{\circ}$ and 3$^{\circ}$ respectively. When account is taken of
these atmospheric emission and absorption effects, the choice of
operating frequencies is $\le 15$\,GHz.

\begin{table}
  \centering
  \begin{tabular}{c | c | p{.55cm} p{.55cm} p{.01cm} p{.55cm} p{.55cm} p{.55cm} p{.55cm} |}
    \hline
    \multirow{2}{*}{$\lambda$} & \multirow{2}{*}{$\nu$}     & \multicolumn{2}{c | }{Zenith $T$}     && \multicolumn{4}{c |}{Total $T$}   \\
    \multirow{2}{*}{(cm)}      & \multirow{2}{*}{(GHz)}     & \multicolumn{2}{c | }{(K)}          && \multicolumn{4}{c |}{(K)}       \\
    \cline{3-4}
    \cline{6-9}
              &                &       &          & & \multicolumn{4}{c}{Elevation} \\
              &                & O$_2$ & H$_2$O   & & 90\celsius{} & 10\celsius{} & 5\celsius{} & 3\celsius{} \\
    \hline
    \hline
    21        & 1.4       & 2.2   & 0.0    && 2.2          & 12.3         & 23          & 41          \\
    5.9       & 5.0       & 2.3   & 0.1    && 2.4          & 14.5         & 29          & 45          \\
    2.9       & 10.0      & 2.4   & 0.2    && 2.6          & 19.5         & 37          & 54          \\
    1.91      & 33        & 7.6   & 3.3    && 10.9         & 75.0         & 140         & 185         \\
    0.33      & 90        & 14.6  & 15.0   && 29.6         & 150          & 240         & 270         \\
    \hline
  \end{tabular}
  \caption{Atmospheric emission as a function of elevation for
    selected frequencies in the range 1--100\,GHz.
    The emission is given
    for 10 mm pwv. Data are from observations
    \citep{Danese1989,Staggs1996,Zhang2012} and from the modelling
    \citep{Pardo2001,Smith1982,Waters1976,Crane1976}.
  }
  \label{tab:atmemissionelevation}
\end{table}

\section{The working system}
\label{sec:proposedsys}

A compact radio telescope suited to the observation of
the Moon in proximity to the Sun is presented in this
section.

\begin{figure}
  \centering
  \includegraphics[width=\imgspace]{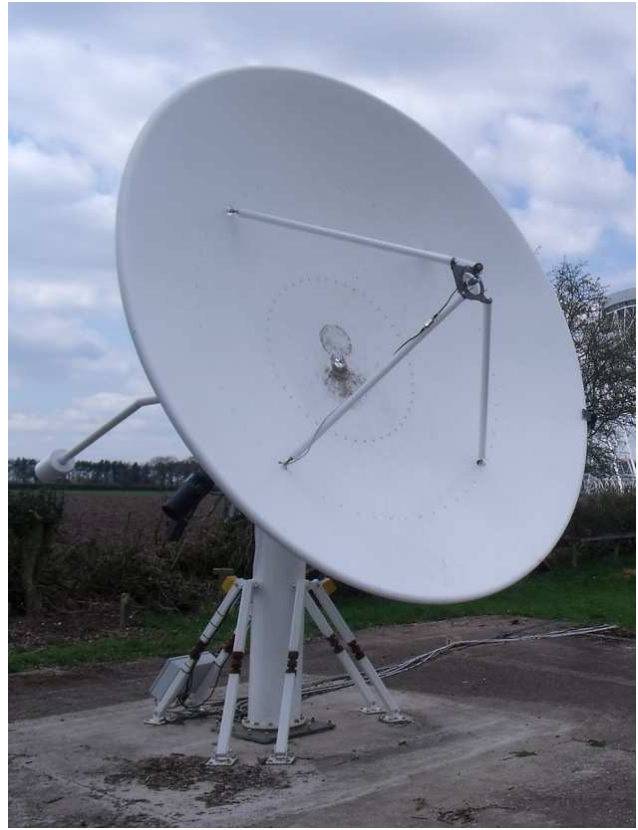}
  \caption{A photograph of the 3.7\,m telescope showing the diagonal
    feed support arrangement
    \label{fig:moontel_outdoor}}
\end{figure}

\subsection{General requirements}
\label{sec:telescope_specification}

\newcommand{\fwhm}{FWHM}
\newcommand{\note}{}

The aim of the telescope is to acquire a set of scans of the sky
temperature where
the Sun and the Moon are located in the 24\,hours prior to and after the time of
New Moon at intervals of $\approx 1$\,hour.
As discussed in the previous sections, the ideal frequency
for such observations
is between 5 and 15\,GHz. In this frequency range, the Sun
brightness is  $\sim 20$\,dB above the
Moon and the atmosphere is still relatively transparent.
We have therefore chosen 10\,GHz as the frequency for this
project. Such frequencies is commonly used for satellite
communications and the components (receivers, detectors,
filters, etc.) are inexpensive and readily available.

In order to achieve its
primary goal, the telescope must have a beamwidth small enough to
resolve the Sun and the Moon even when in close proximity (less than
few degrees). Given that the Moon moves $\approx 30$\,arcmin~per~hour
along the ecliptic with respect to the Sun and their minimum distance
is 0\angdeg{} (as during total solar eclipse), the beamwidth of the
telescope must be $\approx 30$\,arcmin.
These considerations lead to a diameter of the antenna effective aperture
$D \ge \lambda / \theta \approx 4$\,m at the frequency of 10\,GHz
($\lambda = 3$\,cm).

A second
requirement derives from the strong sidelobe level associated with the
much brighter Sun.
The sidelobe level necessary to identify the Moon in the presence of
the Sun, which is $\sim$ 50--100 times brighter at 10\,GHz, is greater
than 25\,dB (a factor of 300) at a projected distance of 1\angdeg{}
to 2\angdeg{} in the sky. This requires a polar diagram for the prime
focus feed which tapers to $\le -15$\,dB at the edge of a parabolic
reflector.
As a result, the required reflector diameter may be larger in order to
accommodate for the feed taper.

\begin{figure*}
  \centering
  \begin{minipage}{125mm}
  \includegraphics[width=\textwidth]{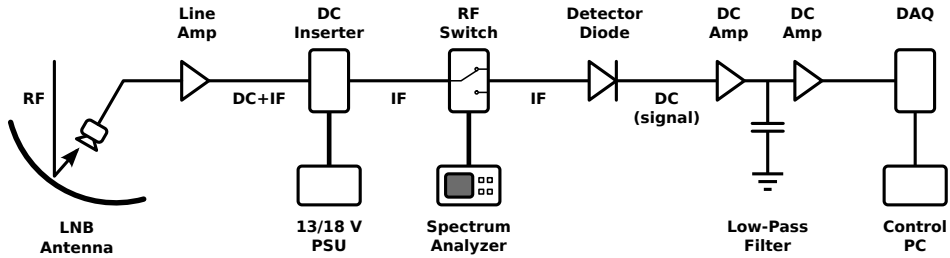}
  \caption{Diagram of the telescope detection system as implemented
  for the New Moon observations.
  }
  \label{fig:detectionchain}
  \end{minipage}
\end{figure*}

The telescope must be agile and have a fast scanning speed in order
to obtain quasi-synchronous scans (snapshot) of the Sun and the Moon
in celestial coordinates and avoid significant relative motion of the Moon
during the observation. For a telescope with \fwhm{}~$\approx 30$\,arcmin
and given the Moon motion with respect to the Sun of
30\,arcmin~per~hour, the scanning time
must be less that 30\,min to avoid significant motion of the Moon.
For an observation performed 24\,hours before or after closest approach, the angular
separation of the Moon and Sun may be as large as $\approx 15\angdeg{}$;
depending on the scanning strategy, to cover an area of
$15 \angdeg{} \! \times 15 \angdeg{}$ in size with a 30\,arcmin
beam leads to a minimum telescope pointing speed of no less than
30\,arcmin~per~sec.

\begin{table}
  \centering
  \begin{tabular}{l | c}
    \hline \textbf{Parameter}  & \textbf{Value}       \\
    \hline
    \hline
    Reflector diameter         & 3.7\,m               \\
    \fwhm{} beamwidth          & 36\,arcmin          \\
    Azimuth drive speed        & $\ge 30$\,arcmin per sec  \\
    Pointing accuracy          & 2\,arcmin         \\
    Frequency                  & 10.8\,GHz            \\
    Bandwidth                  & 1\,GHz               \\
    $T_{sys}$ at zenith        & 100\,K               \\
    \hline
  \end{tabular}
  \caption{Telescope and receiver parameters.}
  \label{tab:proposedsystem}
\end{table}

\subsection{The telescope}

A telescope designed to observe the New Moon has been developed
at the Jodrell Bank Observatory, UK (longitude~$02\angdeg{}18^m$\,W,
latitude~$53\angdeg{}14^m$\,N). The telescope has been
constructed in accordance with the requirements stated in the Sec.~5.1.
The parameters of the telescope are given in Table~\ref{tab:proposedsystem}.

The telescope
receiver is a
commercial room temperature satellite low-noise block downconverter (LNB) with an input
radio frequency (RF) of 10.3\,GHz to 11.3\,GHz.
The LNB contains the low-noise amplifier (LNA), a local oscillator (LO) and a mixer to
down-convert the input frequency to the Intermediate Frequency (IF) of 950\,MHz to
2.0\,GHz.
After a further
20\,dB amplification, the IF~signal is converted into a DC voltage
by a power-law detector diode.
The DC level is then amplified using two operational amplifiers in the
inverting configuration and
filtered using a RC low-pass filter with a
cut-off frequency of 150\,kHz. The two DC amplifiers have $-4.7$
and $-10$ gain respectively.
A commercial data acquisition system (DAQ) containing an
analogue~to~digital
16~bits card measures the filtered DC at a sampling rate of up to~25\,kHz.
Fig.~\ref{fig:detectionchain} shows the telescope detection
system; the diagram includes the LNB power supply (PSU) and an IF switch
used to divert the RF signal to a spectrum analyzer for testing.

The telescope reflector is a parabolic dish 3.7\,m in diameter.
It is illuminated with a horn feed located at the primary focus
to give a beam with sidelobes more
than $20$\,dB lower than the main beam. The feed and receiver system are
supported at the dish focus by four metallic cylindrical legs 7\,cm in
diameter and 1.5\,m in length.
Fig.~\ref{fig:moontel_outdoor} is a photograph of the
telescope; it shows the 3.7\,m reflector, the receiver and the diagonal feed
support arrangement.

The telescope pointing is obtained directly from two 24-bit absolute
rotary encoders on the azimuth and elevation axes. The Sun was used to
test the telescope pointing accuracy and was found to be good to
$\approx 2$\,arcmin using this simple method. This value is
sufficiently small given the large telescope beam. The telescope drive
speed is $\approx 30$\,arcmin~per~second in both azimuth and elevation
directions.

At the frequency of 10.8\,GHz and for a 3.7\,m effective aperture
antenna, the expected main beam \fwhm{} is $\approx 30$\,arcmin.
The beam \fwhm{} was
measured on a geostationary satellite to be $36 \pm 3$\,arcmin. This
value was confirmed by the Moon and Sun measurements (see below).

\subsubsection{Data acquisition and mapping}

The recorded time-ordered-data include the telescope position, the
receiver output
voltage and the time. The UTC time is obtained from an NTP server
using the GPS system as the primary timing source.
A sampling rate of
50--100\,Hz for both the receiver DC voltage and telescope position
gives more than adequate redundancy in the data streams.

The telescope is capable of recording a
$15\angdeg{}\! \times 15\angdeg{}$
map of the sky signal around the Sun and
the Moon every 30--40~minutes at the time of New Moon.
The scanning strategy is to cover the required
sky area with a series
of azimuth scans at intervals of 15\,arcmin in elevation which gives
a sky sampling at the Nyquist rate for a 36\,arcmin beam.
Each 15\angdeg{} azimuth scan takes 20\,sec at the scan rate of
30\,arcmin~per~sec. With a turn-around of 10\,sec to account for
deceleration and reacceleration, the net time for a scan is
30\,sec.
This satisfies the requirement to complete the map of the Sun/Moon
within 30\,min.

A baseline is removed from each scan before it is
incorporated into the final map.
The map is obtained by simple binning of the signal into a two
dimensional array.
Further destriping of the map can be
performed off-line.

The map can be
displayed in celestial coordinates (RA and Dec) for the purpose
of displaying the passage of the Moon past the Sun when determining
the true time of New Moon
\citep{Bruin1977,Ilyas1994,Yallap1998,hoffman2003}.

\subsubsection{Polar diagram measurements}

\begin{figure}
  \centering
  \includegraphics[width=\imgspace]{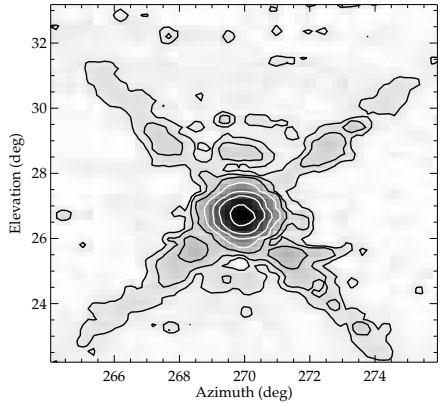}
  \caption{Map of the sidelobe structure on the Sun observed 17:05\,UT
    on 28 May
    2012 at an elevation of 27\angdeg{}. Contours are at 0.1, 0.2, 0.4,
    1.4, 6.3, 20 and 63~per cent of the peak (the lowest 3 contours
    correspond to $-30$, $-27$, $-24$\,dB). Both the circular
    diffraction
    lobe and the diagonal scattering sidelobe structures can be seen
    \label{fig:sunsidelobe}}
\end{figure}

The beam shape is well defined with high signal-to-noise using the
Sun. A map of the sidelobe structure using observations made on 28 May
2012 is shown in Fig.~\ref{fig:sunsidelobe}.
Both the circular inner diffraction sidelobes and the diagonal
scattering sidelobes can be seen.
The diffraction lobe is at a radius of 1\angdeg{}50\angmin{}. An inner
diffraction lobe is suggested at a radius of $\approx 1$\angdeg{} embedded
in the main beam as convolved with the Sun. The diagonal
sidelobes produced by the scattering of Earth and Sun radiation into
the prime beam can be traced out to a radius of 7\angdeg{} from the
beam centre where the amplitude is 0.3\,per cent ($-25$\,dB) of the main
beam response.

\begin{figure}
  \centering
  \includegraphics[width=\imgspace]{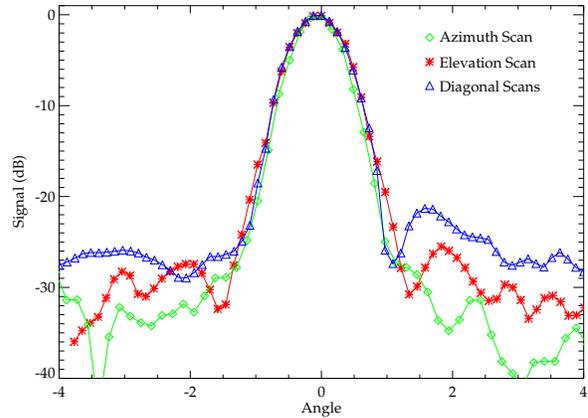}
  \caption{Plot of the beam response showing the main beam and
  sidelobes for the Sun on 28 May 2012 as seen in
  Fig.~\ref{fig:sunsidelobe}. The curves are the azimuth (green
  diamonds) and
  elevation (red stars) scans and the average of the two diagonal
  direction scans (blue triangles).}
  \label{fig:sunsidelobelog}
\end{figure}

The amplitude of the sidelobe structure is quantified
in Fig.~\ref{fig:sunsidelobelog} on a logarithmic dB scale. The
elevation
and azimuth cuts show that the diffraction lobes at a radius of
1\angdeg{}50\angmin{} are $-27$\,dB compared to the peak. The
average diagonal cut shows variations of $-25$ to $-28$\,dB extending from
1\angdeg{}\,30\angmin{} out to 7\angdeg{} radius. Beyond that radius the
level falls further to a level of $\le -30$\,dB (0.1~per cent),
comparable to the noise level at the pixel scale of the map.

It can be seen when comparing observations of the
Sun on different days (Figs.~\ref{fig:sunsidelobe} and
\ref{fig:newmoon}) that the
sidelobes are sufficiently stable and a sidelobe template of the Sun
can be removed from the map. This would more clearly show the Moon and
the Sun as convolved with the 36\,arcmin primary beam without the
confusion of Sun sidelobes.

\subsection{Observation of the Sun and Moon at New Moon on 21 May 2012}

\begin{figure}
  \centering
  \includegraphics[width=\imgspace]{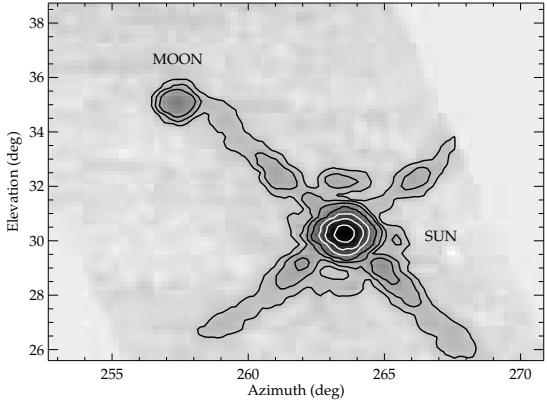}
  \caption{A map of the Sun and Moon
    on 21 May 2012 near the time of
    New Moon. The Sun was observed at 16:27\,UT; the Moon at 16:36\,UT.
    Contours are at 0.1, 0.2, 0.4, 1.4, 6.3, 20, and 63~per cent of
    the Sun peak. The sidelobe levels may be compared directly with
    those in Figure \ref{fig:sunsidelobe} for 28 May 2012 observation.}
  \label{fig:newmoon}
\end{figure}

An observation of the Sun and Moon some 4 hours before sunset on 21
May 2012 is shown in Figure \ref{fig:newmoon}. Contours are given at
the same levels as in Figure \ref{fig:sunsidelobe}. The sidelobe
levels are very similar to those on the 28 May 2012 when the
elevation was similar.
This has implications for the possibility of
developing an algorithm for subtracting a sidelobe template from the
observations in order improve the Moon visibility against the brighter
Sun as described in Sec.~5.2.2. At low elevations ($\le 10$\angdeg{})
account would have to be taken of the greater scattering of Earth
radiation from the feed supports as well as atmospheric absorption and
refraction. These latter effects are calculable and small.

\begin{figure}
  \centering
  \includegraphics[width=\imgspace]{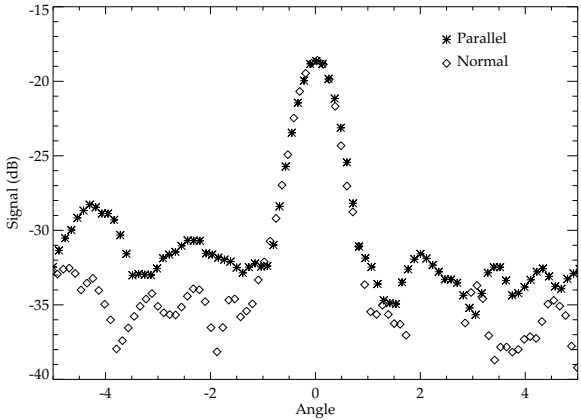}
  \caption{Cuts through the centre of the Moon on 21 May 2012 parallel
  (full line) and perpendicular (dashed line) to the diagonal lobe
  direction. Amplitudes are given as per centage of the
  Sun.}
  \label{fig:moondiacut}
\end{figure}

The Moon is clearly visible even though it lies close to a diagonal
lobe. Cuts across the Moon parallel and perpendicular to the diagonal
lobe are shown in Fig.~\ref{fig:moondiacut}. 
The cut parallel to the sidelobe is $-4$ to $-5$\,dB relative to the
perpendicular cut. The parallel cut shows that the
Moon was 1.37 $\pm$ 0.03~per cent of the Sun, indicating that the Sun was $73.0\pm2.2$
times brighter than the Moon on 21 May 2012. The Moon is clearly detected at about 13\,dB (factor of 20) above the sidelobe structures. The effective noise level in Fig.~\ref{fig:sunsidelobe} is $\approx 0.03$ per cent relative to the Sun.

\subsection{Comparison of Sun to Moon ratio with previous data}

The observed ratio of 73 for the Sun-to-Moon emission at 10.8\,GHz can
be compared with the value predicted from the literature as
outlined in Sections \ref{sec:radiosun}~\&~\ref{sec:radiomoon} and
Tables~\ref{tab:sun}~\&~\ref{tab:moon} of this paper.
The relevant
input data for this calculation of the New Moon on 21 May 2012 are
given on Table~\ref{tab:sunmoonestim}.

\begin{figure*}
  \centering
  \begin{minipage}{125mm}
  \includegraphics[width=\textwidth]{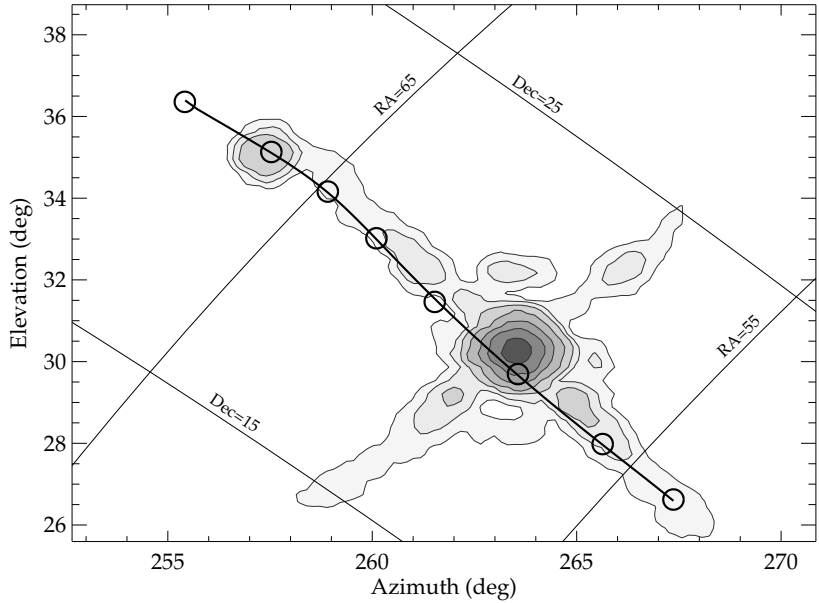}
  \caption{The map of 16\,hr\,36\,min\,UT on 21 May 2012 with
    suporimposed celestial coordinates. The path of the Moon relative
    to the Sun in celestial coordinate is shown at
    4 hour intervals for the previous 24~hours. The Moon passed
    closest to the Sun at 00\,hr\,09\,min\,UT.
  }
  \label{fig:newmoonradec}
  \end{minipage}
\end{figure*}

\begin{table}
\centering
\begin{tabular}{l r c}
  \\
  \hline
  \multicolumn{3}{c}{\bf Sun 21 May 2012}               \\
  \hline
  \hline
  Mean diameter                           & 31.99\,arcmin \\
  Diameter on observation day             & 31.66\,arcmin \\
  Quiet Sun (SS max)                      &                & 16,600\,K   \\
  Quiet Sun (SS min)                      &                & 12,600\,K   \\
  Quiet Sun on 21 May 2012                &                & 14,600\,K   \\
  Sunspot number $R =$ 71                 &                &             \\
  Sunspot emission                        &                & 1,500\,K    \\
  Expected $T_b$                          &                & 16,100\,K   \\
  Sun area correction                     & $-2.1$~per cent       \\
  Corrected T$_b$                         &                & 15,800\,K   \\
  \\
  \hline
  \multicolumn{3}{c}{\bf Moon 21 May 2012}               \\
  \hline
  \hline
  Mean diameter                        & 29.50\,arcmin  \\
  Diameter 21 May 2012                 & 30.09\,arcmin  \\
  Moon mean T$_b$                      &                 & 216\,K \\
  First harminic correction            & $-11$\,K        & 205\,K \\
  Moon area correction                 & +4.0~per cent &  \\
  Corrected T$_b$                      & & 213\,K  \\
  \\
  \hline
  \textbf{Expected ratio Sun/Moon} & $74.2 \pm 7.4$ \\
  \textbf{Observed ratio Sun/Moon} & $73.0 \pm 2.2$ \\
  \hline
  \\
\end{tabular}
\caption{Estimation for the Sun and the Moon brightness temperature at
  10\,GHz on 21 May 2012. The
  Sun and Moon are observed at similar elevations and so there is no
  effect on the Sun/Moon ratio.}
\label{tab:sunmoonestim}
\end{table}

The quiet Sun brightness temperature for 21 May 2012 is taken as the
average of the sunspot maximum and minimum values, namely
14,600\,K. The sunspot emission for a sunspot number $R = 71$ is
1,500\,K at 10\,GHz (Table \ref{tab:sun}). The Sun
diameter on
21 May 2012 (31.66\,arcmin) is less than its mean diameter
(31.99\,arcmin) used for the solar data in
Table~\ref{tab:sun}. This leads to an area correction of
$-2.1$~per cent which gives a reduction of the effective T$_b$ from
16,100\,K to 15,800\,K.

The Moon mean brightness temperature at 10\,GHz is 216\,K with a
correction of $-11$\,K for a first harmonic phase lag of 40\angdeg{} at
the time of New Moon. On correction for the Moon diameter to 21 May
2012, the predicted brightness temperature of the Moon is 213\,K.

The predicted ratio Sun-to-Moon for 21 May 2012 is $74\pm7$. The uncertainty
is mainly in the solar value which conservatively is 10~per cent. This
ratio may be compared with the observed value for that day of $73.0 \pm
2.2$. This agreement confirms that the Sun and Moon parameters
given in Tables~\ref{tab:sun}\&\ref{tab:moon} are a reliable guide for
estimating the relative brightness of the Sun and Moon.

\subsection{Estimation of time of New Moon}

The time of New Moon is defined as the time of minimum elongation of
the Moon from the Sun in celestial coordinates \citep{Bruin1977,hoffman2003,Ilyas1994}.
Figure
\ref{fig:newmoonradec} shows this situation for the New Moon on 21 May
2012 as seen from Jodrell Bank Observatory (longitude 00$^h$09$^m\!$.2,
latitude 53\angdeg{}14\angmin{}). The Moon and Sun positions are for
16\,hr\,36\,min\,UT and 16\,hr\,26\,min\,UT respectively. The movement of
the Moon in celestial coordinates is shown at 4\,hr intervals during
the previous 24 hours.

The plot shows that the Moon past in front of the Sun 16\,hr\,25\,min
previous to the observation shown here, namely at 00\,hr\,12\,min\,UT on
21 May 2012. During this 16\,hr~25\,min interval, the Sun will have moved
+2$^m\!$.5 in RA, thus making the true New Moon at 00\,hr\,09\,min as
seen from Jodrell Bank Observatory. This agrees with the time of an
annular eclipse of the Sun which occurred at 00\,hr00\,min\,UT.

\section{Discussion and conclusions}
\label{sec:conclusion}

\subsection{Review of the aims}

We have shown that it is possible to design a radio telescope for
direct observation of the time of New Moon. Optimal frequencies are in
the range 5--15\,GHz where the ratio of the Sun to Moon is 50--100
($\le$ 20\,dB), so as not to be confused by the sidelobes of the
telescope.

At these frequencies the angular diameter of the Sun approaches the
optical value of 0\angdeg{}$\!$.5. The Moon angular diameter is
0\angdeg{}$\!$.5 at all frequencies. A telescope of moderate size gives a
beamwidth of 0\angdeg{}$\!$.5 which is suitable for clearly separating the
Sun and Moon. The emission from the Earth's atmosphere is manageable
at the low elevation required by the traditional methods of sighting
the New Moon.

\subsection{The radio telescope system}

A system operating at 10.8\,GHz has been built to demonstrate the
practicability of detecting the Moon as it passes the Sun at the time
of New
Moon. The telescope sidelobes are $\approx - 23$\,dB as compared with $ -
18.6$\,dB
for the Moon-to-Sun ratio. These sidelobes are shown to be stable and
can be
accounted for in the analysis by subtracting a beam template of the
Sun sidelobes.

At 10.8\,GHz the New Moon observations can be
followed down to an elevation of $\approx$3\angdeg{} where the
atmospheric emission increases the system noise by $\approx 40$\,K and the
attenuation of the Sun and the Moon signals by $\approx 15$~per cent
(0.6\,dB).

The telescope scanning strategy and pointing accuracy were shown to
give a 15\angdeg{}$\times$15\angdeg{} map of the Moon-Sun field in
30\,min. The aims for the project set out in Sec~5.1 and 6.1 are
achieved.

\subsection{Measurements on 21 May 2012}

The setting Sun-Moon field was observed 16 hours after the annular
eclipse of the Sun at 00\,hr\,00\,min\,UT on 21 May 2012. The Sun/Moon
ratio was $73.0 \pm 2.2$ in
agreement with the data from the literature given in the Tables
\ref{tab:sun} and \ref{tab:moon}. It was found using the observed
images that the Moon can be clearly distinguished from the Sun for any
angular distance above $\approx 2$\angdeg{}. Fig.~\ref{fig:newmoonradec}
is a plot of the path of the Moon relative to the Sun in the RA-Dec
system for the previous 24~hours which shows that the New Moon would
have occurred at 00\,hr\,09\,min as seen from Jodrell Bank Observatory.

\subsection{Conclusions -- the potential of the radio method}

We have made a successful test of a radio method of directly observing
the time of New Moon. It has the advantage over the optical
(naked eye or binocular) method of sighting the New Moon in that at
radio frequencies
(5--15\,GHz) the Moon is $\sim 10^{-2}$ of the brightness of the Sun as
compared with $10^{-5.5}$ for optical frequencies. Furthermore this method
is essentially independent of the weather in the form of water clouds
or atmospheric dust. Observations of the Moon are possible at
elevations down to $\approx 3\angdeg{}$ where the atmospheric opacity is
$\approx 15$\,per cent and the emission is less than the receiver system
temperature at the radio frequency of 10.8\,GHz chosen for this project.

The radio approach can be used at any time of day from sunrise to
sunset.
The present approach is to map the Sun and the Moon field
every 30\,min for 24\,hr either side of the true New Moon.
The resulting ``movies''  will trace the locus of the Moon as it
passes the Sun at 30\,min intervalls which corresponts to
0\angdeg{}$\!.25$ steps. At a single site there is a period of
$12 \pm 2$\,hours when the Moon/Sun system is beneath the
horizon. Accordingly, observations near sunset and sunrise are
important for interpolating the time of closest approach if it occurs
during the night time. The time of true New Moon, when the Moon is nearest to the Sun, is
then directly evident. In principle, this could change the beginning of the lunar calendar month by as much as one day, since it begins when the Moon is observed at the first sunset after conjunction.

\section*{Acknowledgments}

We would like to thank the staff at the Jodrell Bank Observatory for
their assistance during the construction and commissioning of the
telescope. CD acknowledges an STFC Advanced Fellowship, an EU
Marie-Curie IRG
grant under the FP7 and an ERC Starting Grant (no.~307209). We gratefully thank the King Abdulaziz City for Science and Technology (KACST) for funding this project. We would
like to especially thank HH Prince Dr Turki Bin Saud Bin Mohammad
Al-Saud for all his support for this project. Without his support we
could not have completed this work.

\bibliographystyle{mn2e}
\bibliography{yasserpaper}

\end{document}